\documentclass[aps,twocolumn,prd,showpacs,nofootinbib]{revtex4}
\usepackage{amsmath}
\usepackage{graphicx}
\usepackage{dcolumn}
\usepackage{bm}
\usepackage{amssymb}
\usepackage{latexsym}
\usepackage{color}
\usepackage{slashed}

\begin{document}

\title {On Supersymmetric Quantum Spin Model}

\author{R. Kumar$^{a}$}
\email{raviphynuc@gmail.com}
\author{A. Shukla$^{b}$\footnote{Corresponding author.}}
\email{shukla@mail.sysu.edu.cn; ashukla038@gmail.com}
\affiliation {{\it $^{a}$Department of Physics and Astrophysics,
University of Delhi, New Delhi--110007, India}\\
{\it $^{b}$School of Physics and Astronomy,
Sun Yat-Sen University, Guangzhou 510275, China}}

\vskip 5cm

\begin{abstract}
{We discuss various symmetry properties of the ${\cal N} = 2$ supersymmetric quantum spin model in one (0 + 1)-dimension 
of spacetime  and provide their relevance in the realm  of the mathematics of differential geometry. We show {\it one-to-one} 
mapping between the continuous symmetry transformations (and corresponding generators) and de Rham cohomological operators 
of differential geometry. One of the {\it novel} observations is the existence of discrete symmetry transformations which play a 
crucial role in providing the physical realization of the Hodge duality ($\star$) operation. Thus, the present model 
provides a toy model for the Hodge theory.}

\end{abstract}



\maketitle

\section {Introduction}
\label{Sec. 1}
Over the years, it has always been a good idea in theoretical physics to have a  toy model as simple as possible that 
exhibits a certain subtle phenomenon and it is often the case when we study very complicated physical systems. One of 
the prime examples is the study of phase transition in ferromagnetic materials through spin model~\cite{huang}. Spin model 
is extensively studied in field theory, quantum information theory as well as quantum computing and it is one of the 
unifying models which has useful application in different research areas. The two prime examples of spin models are Ising 
and Heisenberg models~\cite{gr,jo}.  It is to be noted that in Ising model there are only two 
possibilities of magnetic spin:  up and down whereas in Heisenberg model, magnetic spin is treated as a quantum variable and it
can be in any random direction. It has been already developed that the spin model is an integrable model which can be exactly solvable 
for the nearest neighbor interactions, long range power law interaction and random interactions~\cite{aa,dm}.

One of the most important features of this model is that it can explain many properties of different kind of physical systems such 
as anti-ferromagnet with competing interaction and infinite ranged spin glasses~\cite{chakr,sachdev,sachdev1}. The spin model also 
plays a pivotal role in studying the finite temperature critical behavior and phase transition~\cite{Sousa,ig}. Motivated by the 
simplicity and its myriad applications, the spin model has been extended to spherical model where the spin of a set of particles on 
lattice interacts with the nearest neighbors with a spherical constraint~\cite{kac,joy,joy1,joy3}. The quantum spherical model has 
led to investigate the phase transitions at finite as well as zero temperatures in ferromagnetic materials and it is also useful to 
study the finite temperature critical behavior of quantum spherical spin-glass~\cite{sch,pt,vojta,singh,tn}. In addition, the quantum 
spherical model  with complex spin variables having continuous local gauge symmetry is shown to be exactly solvable  in thermodynamic 
limit and it exhibits very interesting properties. For instance, the local symmetry is spontaneously broken at finite as well as zero 
temperatures which implies the existence of classical and quantum phase transitions with a non-trivial critical behavior. Furthermore, 
in the continuum limit of the partition function of quantum spherical model with local symmetry is equivalent to the well-known $CP^{n - 1}$ 
model in the limit $n \to \infty $~\cite{gomes}.

Recently, the  supersymmetric version of quantum spherical model (within the framework of superspace formalism) has been discussed 
where {\it two} nilpotent symmetry transformations are found to exist~\cite{gomes1}. Moreover, by using the path integral formalism, 
the critical behavior of this model has been discussed at finite and zero temperatures where the supersymmetric quantum spherical model 
shows critical behavior when the supersymmetry is broken~\cite{das,umezawa,das1}. However, in our present investigation, we focus only 
on the quantum spin model which is obtained from quantum spherical model after relaxing the spherical spin constraint. In this paper, 
we supesymmetrize the quantum spin model and show that there is an one-to-one mapping between symmetry transformations and cohomogical 
operators.

The quantum spin model is one of the simplest models which is physically as well as mathematically very rich. One should note that the 
quantum spherical model has played a key role in explaining the various properties/aspects of the physical systems but this model is 
mathematically less explored especially from the symmetry point of view.  It is to be pointed out that the Ising model of statistical 
mechanics, which is one of the prime examples of quantum spin model, has the interesting mathematical connections with  ${\cal N} = 2$ 
supersymmetry quantum mechanical models and Hodge theory~\cite{vafa}.

In the recent spate of papers~\cite{rpm1,rpm2,rpm3,rpm4}, in addition to the usual (anti-)BRST transformations, the nilpotent 
(anti-)co-BRST  transformations also  exist for the Abelian $p$-form $(p = 1, 2, 3)$ gauge theories in $D = 2p$-dimensions of 
spacetime within the framework of BRST formalism. As far as the models of Hodge theory are concerned, within the framework of 
BRST formalism, Abelian $p$-form (i.e. $p =1, 2,3$) gauge theories have been shown to be the tractable  field-theoretic model 
for the Hodge theory in $D = 2p$-dimensions of spacetime where all the de Rham cohomological operators of differential geometry 
find their physical realizations in terms of the continuous symmetry transformations (and their corresponding generators) and 
the discrete symmetry of the theory provides the analogue of Hodge duality operation.  In addition, $(0 + 1)$-dimensional 
toy models (e.g. rigid rotor and Christ-Lee model) are also shown to be the models for Hodge theory ~\cite{rpm5,ra}. By exploiting 
the supervariable technique, the supersymmetric quantum mechanical models with harmonic and general potentials have also been 
proven to be the model for Hodge theory~\cite{rm,rm1,kam,as}. It is important to mention that the (anti-)BRST and supersymmetric 
transformations are nilpotent of order  two but BRST and anti-BRST  transformations anticommute  whereas the anticommutator of two 
supersymmetric transformations is {\it non-zero} and equivalent to the time-translation.

Our present investigation is motivated by the following factors. First, we explore the various continuous  symmetries posses by our 
present model. In fact, we show that in addition to the supersymmetric transformations ($s_1, s_2$), there also exists one more 
continuous transformation ($s_w$) under which Lagrangian for supersymmetric quantum spin model remains invariant. The latter symmetry 
transformation $s_w$ is bosonic in nature. Second, we prove that nilpotent supersymmetric  transformations ($s_1$, $s_2$) and bosonic 
transformation ($s_w$) obey a similar algebra as satisfied by the de Rham cohomological operators of differential geometry. Third, we 
also show the existence of novel discrete symmetry transformations under which the Lagrangian for the supersymmetric quantum spin model  
remains invariant. These bosonic and discrete transformations were {\it not} reported in \cite{gomes1}.  It is important to mention here 
that these discrete symmetry transformations play very important role because these transformations provides physical realization of  
Hodge duality ($\star$). Finally, we show that the ${\cal N} = 2$ supersymmetric quantum spin model provides a simple toy model for the 
Hodge theory in its own right.

The outline of our present investigation is as follows. In Sec. \ref{Sec. 2}, we discuss very brief about the quantum spin model. 
Sec. \ref{Sec. 3} deals with the ${\cal N} = 2$ supersymmetrization  of the quantum spin model. The various continuous symmetries 
(and corresponding charges) and discrete symmetries are discussed in Secs. \ref{Sec. 4} and \ref{Sec. 5}, respectively. We capture 
the physical realization of the abstract de Rham cohomological operators of differential geometry in terms of the symmetry properties 
of the present model in Sec. \ref{Sec. 6}. The cohomological aspects are discussed in this section, too.  Finally, in Sec. \ref{Sec. 7}, 
we provide the concluding remarks.

In  Appendix \ref{Appendix A}, we discuss about the on-shell nilpotent symmetries for the supersymmetric quantum spin model.

\section{Quantum spin model: a brief introduction}
\label{Sec. 2}
The quantum version of the classical spin model is described by the following Hamiltonian:
\begin{eqnarray}
H_0 = \frac{g}{2}\, \sum_{\bf r}\, P^2_{\bf r} + \frac{1}{2}\, \sum_{{\bf r}, {\bf r'}} J_{{\bf r}, {\bf r'}}\, S_{\bf r}\, S_{\bf r'} 
+ h \sum_{\bf r} S_{\bf r}, \label{ham0}
\end{eqnarray}
where $\{S_{\bf r}\}$ is a set of continuous spin variables associated to each lattice site  ${\bf r}$ on a $d$-dimensional hypercubic 
lattice and ranging from $-\infty$ to $+ \infty$, the interaction energy $J_{{\bf r}, {\bf r'}} = J(|{\bf r} - {\bf r'}|)$ depends upon 
the distance between the lattice sites ${\bf r}$ and ${\bf r'}$ where the interaction energy is assume to be translationally invariant. 
Here $P_{\bf r}$ is the canonical conjugate momentum corresponding to the dynamical spin variable $S_{\bf r}$,  $h$ is an external field 
and $g$ defines the quantum coupling constant. In the limit $g \to 0$, the quantum spin model reduces to the classical spin model.

Using the Legendre transformations and switching-off external field (i.e. $h = 0$), we obtain the following Lagrangian   
\begin{eqnarray}
L_0 = \frac{1}{2g}\, \sum_{\bf r}\, \dot{S}^2_{\bf r} 
- \frac{1}{2}\, \sum_{{\bf r}, {\bf r'}} J_{{\bf r}, {\bf r'}}\, S_{\bf r}\, S_{\bf r'},
\label{lag0}
\end{eqnarray}
where ${\dot S}_{\bf r} = \frac{d}{dt} S_{\bf r}$ is the generalized velocity  in the configuration space. The canonical 
conjugate variables $S_{\bf r}$ and $P_{\bf r}$ obey the following commutations relations (with $\hbar = 1$):
\begin{eqnarray}
\big[S_{\bf r}, \, P_{\bf r'} \big] = i\, \delta_{{\bf r},{\bf r'}}, 
\quad  \big[S_{\bf r}, \, S_{\bf r'} \big] = 0, \quad \big[P_{\bf r}, \, P_{\bf r'} \big] = 0, 
\label{commu}
\end{eqnarray}
where $P_{\bf r} = \frac{1}{g}\,{\dot S}_{\bf r}$ is the canonical conjugate momentum  corresponding to
the dynamical variable  $S_{\bf r}$.

In our next sections, we supersymmetrize the above Lagrangian by means of ${\cal N} = 2$ superspace formalism
and discuss the various continuous and discrete symmetries.

\section{${\cal N} = 2$ supersymmetrization of quantum spin model}
\label{Sec. 3}
We consider the  ${\cal N} = 2$ real supervariable $\Phi_{\bf r}(t, \theta, \bar \theta)$ associated to each lattice site ${\bf r}$ 
defined on $(1,2)$-dimensional supermanifold. The $(1,2)$-dimensional superspace is parameterized by the superspace coordinates 
$(t, \theta, \bar \theta)$. Here $t$ is the bosonic time-evolution parameter and $\theta, \bar \theta$ are the Grassmannian variables 
(with $\theta^2 = \bar \theta^2 = 0, \; \theta \bar \theta + \bar \theta \theta = 0$). The supervariable 
$\Phi_{\bf r}(t, \theta, \bar \theta)$ can be expanded along the Grassmannian directions as:
\begin{eqnarray}
\Phi_{\bf r}(t, \theta, \bar \theta) = S_{\bf r}(t) + i \theta \bar \psi_{\bf r}(t) 
+ i \bar \theta \psi_{\bf r}(t) + \theta \bar \theta F_{\bf r}(t),
\label{superv}
\end{eqnarray}
where $S_{\bf r}$, $\psi_{\bf r}$, $\bar \psi_{\bf r}$ are the  basic dynamical  variables and $F_{\bf r}$ is an auxiliary variable 
for our ${\cal N} = 2$ SUSY quantum spherical system. The fermionic variables $\psi_{\bf r}$, $\bar \psi_{\bf r}$ 
(with $\psi^2_{\bf r} =  \bar \psi^2_{\bf r} = 0,\, \psi_{\bf r}  \bar \psi_{\bf r}  + \bar \psi_{\bf r} \psi_{\bf r}  = 0 $) are 
the ${\cal N} =2$ supersymmeric counterparts of the bosonic variable $S_{\bf r}$. All the basic and auxiliary variables are 
the function of time-evolution parameter $t$ only.

Now we introduce two fermionic supercharges $Q$ and $\bar Q$ (with $Q^2 = 0, \; \bar Q^2 = 0$) for the above ${\cal N} = 2$ 
supersymmetric quantum spin model as
\begin{eqnarray}
Q = \frac{\partial}{\partial \bar \theta} + i\,\theta\,\frac{\partial}{\partial t}, \qquad \qquad
\bar Q = \frac{\partial}{\partial  \theta} + i\, \bar \theta\,\frac{\partial}{\partial t},
\end{eqnarray}
where the partial derivatives $\partial/\partial t,\; \partial/\partial \theta,\; \partial/\partial \bar \theta$
are defined on the $(1, 2)$-dimensional supermanifold. 
The above fermionic charges  turn out to be the generators of the following translations  in the superspace:
\begin{eqnarray}
&&t\longrightarrow t' = t + i \,(\epsilon\, \bar \theta + \bar \epsilon\, \theta), \nonumber\\
&& \theta \longrightarrow \theta' = \theta + \epsilon, \qquad  \bar \theta \longrightarrow \bar \theta' = \bar \theta + \bar \epsilon,
\end{eqnarray}
where $\epsilon$ and $\bar \epsilon$ (with $\epsilon^2 = 0$, $\bar \epsilon^2 = 0$, $\epsilon\, \bar \epsilon 
+ \bar \epsilon\, \epsilon = 0$) are the global and infinitesimal shift transformation parameters along the Grassmannian 
directions on the $(1, 2)$-dimensional supermanifold.

The supersymmetric transformation $(\delta)$ on the supervariable can be expressed in terms of the supercharges 
$Q$ and $\bar Q$ as illustrated below
\begin{eqnarray}
\delta \Phi_{\bf r}(t, \theta, \bar \theta) &=& \delta S_{\bf r}(t) + i\theta\, \delta\bar \psi_{\bf r}(t) + i \theta\, \delta \psi_{\bf r}(t) 
+ \theta \bar \theta\, \delta F_{\bf r}(t) \nonumber\\
&=& (\bar \epsilon\, Q + \epsilon \, \bar Q)\, \Phi_{\bf r}(t, \theta, \bar \theta) \nonumber\\
&\equiv& \Phi_{\bf r}(t', \theta', \bar \theta') - \Phi_{\bf r}(t, \theta, \bar \theta)
\end{eqnarray}
The transformation $(\delta)$ can be divided   into two infinitesimal transformations $\delta_1$ and $\delta_2$ because 
of the presence of ${\cal N} = 2$ supersymmetry. These are listed as follows:
\begin{eqnarray}
&& \delta_1 S_{\bf r} = i\, \bar \epsilon\,  \psi_{\bf r},   \qquad \delta_1 \bar \psi_{\bf r} = - \bar \epsilon\,(\dot S_{\bf r} + i\, F_{\bf r}), \nonumber\\
&& \delta_1 F_{\bf r} =  - \bar \epsilon\,\dot \psi_{\bf r},	\qquad \delta_1 \psi_{\bf r} = 0,\nonumber\\
&& \nonumber\\
&& \delta_2 S_{\bf r} = i\, \epsilon \, \bar \psi_{\bf r}, \qquad \delta_2  \psi_{\bf r} = -  \epsilon\, (\dot S_{\bf r} - i\, F_{\bf r}), \nonumber\\
&& \delta_2 F_{\bf r} = \epsilon\,\dot {\bar \psi}_{\bf r}, \qquad \delta_2 \bar  \psi_{\bf r} = 0,
\label{susy1}
\end{eqnarray}
where we have defined $\dot S_{\bf r} = d S_{\bf r}/dt$, $\dot \psi_{\bf r} = d\psi_{\bf r}/dt$, $\dot {\bar \psi}_{\bf r} = d \bar \psi_{\bf r}/dt.$
It can be readily checked that $\delta_1$ and $\delta_2$ are off-shell nilpotent of order two 
(i.e. $\delta^2_1 = 0, \; \delta^2_2 = 0$).

With the help of super-covariant derivatives ${\cal D}$ and $\bar{\cal D}$ defined as   
\begin{eqnarray}
{\cal D} = \frac{\partial}{\partial \bar \theta} - i\,\theta\,\frac{\partial}{\partial t}, \qquad \quad
\bar {\cal D} = \frac{\partial}{\partial  \theta} - i\, \bar \theta\,\frac{\partial}{\partial t},
\end{eqnarray}
we can write the general Lagrangian for the ${\cal N} = 2$ supersymmetric quantum spin model in the following fashion  
\begin{eqnarray}
L &=& \int d\theta \,d\bar \theta \bigg[\frac{1}{2g}\,\sum_{\bf r} {\cal D} \Phi_{\bf r}(t, \theta, \bar \theta)\, 
\bar {\cal D} \Phi_{\bf r}(t, \theta, \bar \theta) \nonumber\\
&&  +\, \frac{1}{2 \sqrt{g}}\,\sum_{{\bf r},{\bf r}'} U_{{\bf r}, {\bf r}'}\,\Phi_{\bf r}(t, \theta, \bar \theta)\,\Phi_{{\bf r}'}(t, \theta, \bar \theta) \bigg].
\end{eqnarray}
After performing  the Grassmannian integrations,  we obtain the desired Lagrangian for the ${\cal N} = 2$ supersymmetric quantum spin model: 
\begin{eqnarray}
L &=& \frac{1}{2g}\,\sum_{\bf r} {\dot S}_{\bf r}^2 + \frac{i}{g} \sum_{\bf r} \bar \psi_{\bf r} \, \dot \psi_{\bf r} 
+ \frac{1}{2g}\,\sum_{\bf r} F^2_{\bf r} \nonumber\\
&-& \frac{1}{\sqrt{g}} \sum_{{\bf r},{\bf r}'} U_{{\bf r}, {\bf r}'} \,S_{\bf r} \,F_{{\bf r}'} 
 - \frac{1}{\sqrt{g}} \sum_{{\bf r},{\bf r}'} U_{{\bf r}, {\bf r}'}\, \bar \psi_{\bf r} \,\psi_{{\bf r}'}.
\label{lag1}
\end{eqnarray}
In the above, we can re-scale the dynamic fermionic variables  $\psi_{\bf r} \to g^{1/4} \Psi_{\bf r}$,   $\bar \psi_{\bf r} \to g^{1/4} \bar \Psi_{\bf r}$,
$\epsilon \to g^{1/4} \varepsilon$ as well as global Grassmannian parameters  $\bar \epsilon \to g^{1/4} \bar \varepsilon$ without any loss of generality. 
As a consequence, the  Lagrangian (\ref{lag1}) for the supersymmetric quantum spin model takes the following form
\begin{eqnarray}
L &=& \frac{1}{2g}\,\sum_{\bf r} {\dot S}_{\bf r}^2 + \frac{i}{\sqrt{g}} \sum_{\bf r} \bar \Psi_{\bf r} \, \dot \Psi_{\bf r} 
+ \frac{1}{2g}\, \sum_{\bf r} F^2_{\bf r} \nonumber\\
&-& \frac{1}{\sqrt{g}} \sum_{{\bf r},{\bf r}'} U_{{\bf r}, {\bf r}'} \,S_{\bf r} \,F_{{\bf r}'} 
- \sum_{{\bf r},{\bf r}'} U_{{\bf r}, {\bf r}'}\, \bar \Psi_{\bf r} \,\Psi_{{\bf r}'}
\label{lag2}
\end{eqnarray}
and the supersymmetric transformations (\ref{susy1}) read
\begin{eqnarray}
&& \delta_1 S_{\bf r} = i\, \bar \varepsilon\, \sqrt{g} \,\Psi_{\bf r},   
\qquad \delta_1 \bar \Psi_{\bf r} = - \bar \varepsilon\,(\dot S_{\bf r} + i\, F_{\bf r}), \nonumber\\
&& \delta_1 F_{\bf r} =  - \bar \varepsilon\, \sqrt{g}\, \dot \Psi_{\bf r},	\qquad \delta_1 \Psi_{\bf r} = 0,\nonumber\\
&& \nonumber\\
&& \delta_2 S_{\bf r} = i\, \varepsilon \,\sqrt{g} \bar \Psi_{\bf r}, 
\qquad \delta_2  \Psi_{\bf r} = -  \varepsilon\, (\dot S_{\bf r} - i\, F_{\bf r}), \nonumber\\
&& \delta_2 F_{\bf r} = \varepsilon\, \sqrt{g}\,\dot {\bar \Psi}_{\bf r}, \qquad \delta_2 \bar  \Psi_{\bf r} = 0.
\label{16}
\end{eqnarray}
For our further discussions, we shall work with the above Lagrangian (\ref{lag2}).

By exploiting the Legendre transformations, the Hamiltonian for the above system is given by
\begin{eqnarray}
H &=& \frac{g}{2}\, \sum_{\bf r} P^2_{\bf r} 
+\frac{1}{\sqrt{g}} \sum_{{\bf r}, {\bf r}'} U_{{\bf r},{\bf r}'}\, S_{\bf r} \, F_{{\bf r}'} - \frac{1}{2g}\, \sum_{\bf r} F^2_{\bf r} \nonumber\\
&& +\, \sum_{{\bf r}, {\bf r}'} U_{{\bf r},{\bf r}'}\, \bar \Psi_{\bf r} \, \Psi_{{\bf r}'}.
\label{17}
\end{eqnarray}  
The canonical conjugate momenta $P_{\bf r} = \frac{1}{g}\, \dot S_{\bf r}$ and $\Pi_{\bf r} = - \frac{i}{\sqrt{g}}\, \bar \Psi_{\bf r}$
(derived from Lagrangian (\ref{lag2})) corresponding to the dynamical variables $S_{\bf r}$ and $\Psi_{\bf r}$, respectively, obey the 
following non-vanishing canonical (anti)commutation relations: 
\begin{eqnarray}
&& [S_{\bf r}, \,P_{{\bf r}'}] = i \,\delta_{{\bf r}, {\bf r}'}, \nonumber\\
&& \{\Psi_{\bf r}, \, \Pi_{\bf r'}\} = - \frac{i}{\sqrt g} \,\delta_{{\bf r}, {\bf r}'} \Rightarrow 
\{\Psi_{\bf r}, \bar \Psi_{{\bf r}'} \} = \delta_{{\bf r}, {\bf r}'}, \qquad
\end{eqnarray}
where rest of the (anti)commutators are turn out to be zero.

\section{Continuous symmetries and conserved charges}
\label{Sec. 4}
In this section, we discuss various continuous symmetries of the present model. The Lagrangian respects the following off-shell 
nilpotent (i.e. $s^2_1 = 0$ and $s^2_2 = 0$)  ${\cal N} = 2$ supersymmetric transformations:
\begin{eqnarray}
&& s_1 S_{\bf r} = i\, \sqrt{g} \,\Psi_{\bf r},   \qquad s_1 \bar \Psi_{\bf r} = - (\dot S_{\bf r} + i\, F_{\bf r}), \nonumber\\
&&s_1 F_{\bf r} =  - \sqrt{g}\, \dot \Psi_{\bf r},	\qquad s_1 \Psi_{\bf r} = 0,\nonumber\\
&& \nonumber\\
&& s_2 S_{\bf r} = i\, \sqrt{g} \bar \Psi_{\bf r}, \qquad s_2  \Psi_{\bf r} = -  (\dot S_{\bf r} - i\, F_{\bf r}), \nonumber\\
&&s_2 F_{\bf r} = \sqrt{g}\,\dot {\bar \Psi}_{\bf r}, \qquad s_2 \bar  \Psi_{\bf r} = 0,
\label{20}
\end{eqnarray}
where  we have captured the global Grassmannian parameters $\varepsilon$ and $\bar \varepsilon$ in the supersymmetric 
transformations $s_1$ and $s_2$. For the sake brevity,  we have chosen:  $\delta_1 = \bar \varepsilon \,s_1$ and 
$\delta_2 = \varepsilon \,s_2$ (cf. (\ref{16})). Under these off-shell nilpotent and continuous supersymmetric 
transformations, the Lagrangian remains quasi-invariant (i.e. transforms to a total time derivative) as:  
\begin{eqnarray}
&& {\hspace{-1cm}} s_1 L =\frac{d}{dt}\Bigg[\sum_{{\bf r},{\bf r}'} U_{{\bf r},{\bf r}'}\, S_{\bf r} \Psi_{{\bf r}'} \Bigg], \nonumber\\
&& {\hspace{-1cm}} s_2 L = \frac{d}{dt}\Bigg[\frac{1}{\sqrt{g}} \sum_{\bf r} \bar \Psi_{\bf r} \Big(i {\dot S}_{\bf r} + F_{\bf r} \Big) 
- \sum_{{\bf r},{\bf r}'} U_{{\bf r},{\bf r}'} \bar \Psi_{{\bf r}}S_{{\bf r}'} \Bigg].  
\end{eqnarray}
As a consequence, the action integral ($\int dt L $) remains invariant (i.e. $s_1 \int dt L  = 0$ and $s_2 \int dt L  = 0$).

It is clear that even though continuous transformations $s_1$ and $s_2$ are off-shell nilpotent of order two (i.e. $s^2_1  = 0$ and 
$s^2_2 = 0$) but they do not anticommute  i.e. $\{s_1, s_2 \} \ne 0$ for any generic variable. In fact, the anticommutator leads to 
the time translation (modulo a constant factor). This is one of the characteristic features of the ${\cal N} = 2$ supersymmetric models. 
As a consequence, we define a continuous bosonic symmetry $s_w =\{s_1, s_2 \} $. For any generic variable $\chi$, we obtain the bosonic 
symmetry transformation:   
\begin{eqnarray}
s_w \,\chi = - 2i \sqrt{g} \dot \chi, \qquad \chi(t) = S_{\bf r}, \, \Psi_{\bf r}, \, \bar \Psi_{\bf r}, \, F_{\bf r}, 
\end{eqnarray}
under which  the Lagrangian transforms to a total time derivative as
\begin{eqnarray}
s_w L = \frac{d}{dt} \left(-2i \sqrt{g} L \right).  
\end{eqnarray}
Thus, the action integral remains invariant under bosonic symmetry transformation.

According to Noether's theorem, the invariance of the action under the continuous symmetry transformations $s_1$, $s_2$ and $s_w$ 
lead to the conserved charges. These are listed as follows:
\begin{eqnarray}
Q &=& \frac{1}{\sqrt{g}} \sum_{\bf r}\big(i \dot S_{\bf r} - F_{\bf r} \big) \Psi_{\bf r}, \nonumber\\
\bar Q &=& \frac{1}{\sqrt{g}} \sum_{\bf r} \bar \Psi_{\bf r} \big(i \dot S_{\bf r} + F_{\bf r} \big), \nonumber\\
Q_w &=&  -2i \sqrt{g}\Bigg(\frac{1}{2}\, \sum_{\bf r} \dot S^2_{\bf r} 
+\frac{1}{\sqrt{g}} \sum_{{\bf r}, {\bf r}'} U_{{\bf r},{\bf r}'}\, S_{\bf r} \, F_{{\bf r}'} \nonumber\\
&& -\, \frac{1}{2g}\, \sum_{\bf r} F^2_{\bf r}  + \sum_{{\bf r}, {\bf r}'} U_{{\bf r},{\bf r}'}\, \bar \Psi_{\bf r} \, \Psi_{{\bf r}'}\Bigg)\nonumber\\
&\equiv& -2i\sqrt{g} \bigg(\frac{g}{2}\, \sum_{\bf r} P^2_{\bf r} 
+\frac{1}{\sqrt{g}} \sum_{{\bf r}, {\bf r}'} U_{{\bf r},{\bf r}'}\, S_{\bf r} \, F_{{\bf r}'} \nonumber\\
&& -\, \frac{1}{2g}\, \sum_{\bf r} F^2_{\bf r} + \sum_{{\bf r}, {\bf r}'} U_{{\bf r},{\bf r}'}\, \bar \Psi_{\bf r} \, \Psi_{{\bf r}'}\Bigg) \nonumber\\
&=& -2i\sqrt{g} \,H. \qquad
\end{eqnarray}
It is to be noted that the bosonic charge $Q_w$ is nothing but the Hamiltonian of system (multiplied by a constant factor: $-2i {\sqrt g}$).
The conservation of the above charges can be proven by exploiting the following Euler-Lagrange equations of motion:
\begin{eqnarray}
&& \ddot S_{\bf r} + \sqrt{g}\, \sum_{{\bf r}'} U_{{\bf r}, {\bf r}'}\, F_{{\bf r}'} = 0, \nonumber\\
&& F_{\bf r} = \sqrt{g} \sum_{{\bf r}'} U_{{\bf r},{\bf r}'}\,S_{\bf r'},\nonumber\\
&& \dot \Psi_{\bf r} + i \sqrt{g}\, \sum_{{\bf r}'} U_{{\bf r}, {\bf r}'}\, \Psi_{{\bf r}'} = 0, \nonumber\\
&& \dot {\bar \Psi}_{\bf r} - i \sqrt{g} \,\sum_{{\bf r}'} U_{{\bf r}, {\bf r}'}\,\bar \Psi_{{\bf r}'} = 0,
\label{25}
\end{eqnarray}
which have been derived from the Lagrangian (\ref{lag2}).

\section{Discrete Symmetries}
\label{Sec. 5}
In addition to the above continuous symmetries, the Lagrangian (\ref{lag2}) also respects the following discrete 
symmetry transformations  
\begin{eqnarray}
&& t \to - t, \qquad S_{\bf r} \to \mp S_{\bf r}, \quad \Psi_{\bf r} \to \pm \bar \Psi_{\bf r}, \nonumber\\
&& \bar \Psi_{\bf r} \to \mp  \Psi_{\bf r},  \qquad  F_{\bf r} \to \mp F_{\bf r}. 
\label{discr1}
\end{eqnarray}
It is to be noted that we have time-reversal symmetry (i.e. $t \to - t$) in the theory. In addition to the above discrete 
symmetry, we also have another set of discrete symmetry transformations 
\begin{eqnarray}
&& t \to - t, \quad S_{\bf r} \to \pm S_{\bf r}, \quad \Psi_{\bf r} \to \pm i \bar \Psi_{\bf r}, \nonumber\\
&& \bar \Psi_{\bf r} \to \pm i  \Psi_{\bf r}, \qquad  F_{\bf r} \to \pm F_{\bf r},   
\label{discr2}
\end{eqnarray}
which leave the Lagrangian invariant. Under both set of discrete symmetry transformations the conserved charges transform 
in the following fashion:  
\begin{eqnarray}
& *\, Q = + \,\bar Q , \qquad * \,\bar Q = -\,Q,& \nonumber\\
& * \,Q_w = +\, Q_w, \qquad * \,H = +\, H, &
\end{eqnarray}
where $*$ corresponds to the discrete transformations (\ref{discr1}) and (\ref{discr2}). It is clear from the above that under 
the discrete symmetry transformations $Q \to +	\bar Q$ and $\bar Q \to - Q$. This is like the case of electromagnetic duality 
for the source free Maxwell's theory where ${\bf E} \to + {\bf B}$ and  ${\bf B} \to -{\bf E}$.  We note that the two successive 
operation of discrete transformations on any generic variable yields
\begin{eqnarray}
&& *(* S_{\bf r}) = +\, S_{\bf r}, \qquad *(* F_{\bf r}) = +\, F_{\bf r}, \nonumber\\
&& *(* \Psi_{\bf r}) = -\, \Psi_{\bf r}, \qquad *(* \bar \Psi_{\bf r}) = - \,\bar \Psi_{\bf r}.
\label{29}
\end{eqnarray}
Similarly, under the two successive transformations (\ref{discr1}) and (\ref{discr2}), the conserved charges  transform  
as: $*(*Q) = - Q$,  $*(*\bar Q) = - \bar Q$,
 $*(*Q_w) = + Q_w$ and  $*(*H) = + H$. 
Furthermore, the interplay of fermionic and discrete symmetry transformations provides us an interesting relationship: 
\begin{eqnarray}
s_2 \chi = \pm * s_1 * \chi, 
\label{30}
\end{eqnarray}
where $(\pm)$ signs in the r.h.s. of the above expression correspond to the generic variable being (bosonic) fermionic in nature. 
In fact, the $(\pm)$ signs are dictated by two successive operations of discrete transformations on the generic variable (cf. (\ref{29})).
Furthermore,  for any generic variable there also exists a reverse relationship (i.e. $s_1 \chi = \mp * s_2 *\chi$) corresponding to the
above relationship (\ref{30}).

\section{Symmetry algebra and cohomological aspects}
\label{Sec. 6}
In our previous sections, we have discussed two fermionic symmetries and a bosonic symmetry. The operator form of the 
continuous and discrete symmetry transformations obey the following interesting algebra\cite{egh,vh,dght}:     
\begin{eqnarray}
& s^2_1 = 0, \quad s^2_2 = 0, \quad \{s_1, \; s_2\} = s_w,&\nonumber\\
& [s_w, \; s_1] = 0, \quad [s_w, \; s_2] = 0, \quad s_2 = \pm * s_1*. &
\end{eqnarray}
The above algebra among the continuous symmetry transformations is reminiscent of the algebra obeyed by the de Rham cohomological 
operators of differential geometry. The latter algebra is given as follows\cite{egh,vh,dght,smm}: 
\begin{eqnarray}
& d^2 = 0, \qquad \delta^2 = 0, \qquad \{d, \; \delta\} = \Delta,&\nonumber\\
& [\Delta, \; d] = 0, \qquad [\Delta, \; \delta] = 0, \qquad \delta = \pm \star d \star,& 
\label{32}
\end{eqnarray}  
where $d$, $\delta$, $\Delta$ and $\star$ are the exterior derivative, co-exterior derivative, Laplacian operator and Hodge duality 
operation defined on a D-dimensional compact manifold without boundary. The $(\pm)$ signs in the last relation depend on the 
dimensionality of the space as well as degree of a given form. It is clear that there is {\it one-to-one} mapping between the 
continuous transformations $(s_1, s_2, s_w)$ and de Rham cohomological operators $(d, \delta, \Delta)$. In fact, one can identify 
exterior derivative $d$ with fermionic transformation $s_1$, co-exterior derivative $\delta$ with $s_2$ and Laplacian operator 
$\Delta$ with bosonic transformation $s_w$. The discrete symmetry transformations $*$ in (\ref{discr1}) and (\ref{discr2}) provide 
the analogue of Hodge duality $\star$ operation.

As far as the properties of the differential operators are concerned, we note that the exterior derivative, when acts on a given
form $f_n$, raises the degree of the form by one (i.e. $d f_n \sim f_{n+1}$) where $n$ is the degree of form whereas co-exterior 
derivative decreases the degree by one (i.e. $\delta f_n \sim f_{n-1}$). The Laplacian operator does not effect the degrees
of the form (i.e. $\Delta f_n \sim f_n$). We shall captures these properties in terms of the conserved charges.

Before going into details, we first point out that the  conserved charges corresponding to the continuous 
symmetry transformations obey the standard ${\cal N} = 2$ supersymmetric algebra:
\begin{eqnarray}
&& Q^2 = 0, \quad \bar Q^2 = 0, \quad \big\{Q, \; \bar Q\big \} = Q_w = -2i \sqrt{g}\,H, \nonumber\\
&& \big[H, \; Q \big] = 0, \qquad \big[H, \; \bar Q \big] = 0, 
\label{33}
\end{eqnarray}
The above algebra is exactly similar to the algebra followed by the de Rham cohomological operators (cf. (\ref{32})).
It is clear that $Q_w$ is like a Casimir operator of the above algebra. In other words, one can also say that the Hamiltonian 
of the theory is the Casimir operator. Due to the validity of last two relationships in (\ref{33}) and for the non-singular 
Hamiltonian, we equivalently have $\big[H^{-1},\; Q \big] = 0$ and $\big[H^{-1}, \; \bar Q \big] = 0$. As a result, $H^{-1}$ 
would also be the Casimir operator for the above algebra.

Form the above arguments, the following algebraic relations among the conserved charges are true:   
\begin{eqnarray}
&& \left[\frac{Q \bar Q}{H}, \; Q \right] = +Q, \qquad \left[\frac{Q \bar Q}{H}, \; \bar Q \right] = - \bar Q, \nonumber\\
&& \left[\frac{\bar Q  Q}{H}, \; Q \right] = - Q, \qquad \left[\frac{\bar Q Q}{H}, \; \bar Q \right] = + \bar Q,  \nonumber\\
&& \left[\frac{Q \bar Q}{H}, \; H \right] = 0, \qquad \quad  \left[\frac{\bar Q Q}{H}, \; H \right] = 0.
\label{34}
\end{eqnarray}
In view of the above algebra, let us define an eigenstate $|\alpha \rangle_p$ with respect to an operator $\frac{Q \bar Q}{H}$  
having eigenvalue $p$ such that $\frac{Q \bar Q}{H} |\alpha \rangle_p = p |\alpha \rangle_p$. We further point out 
$\left(\frac{Q \bar Q}{H}\right)^\dagger = \frac{Q \bar Q}{H}$ and it is a fermion number operator \cite{salo,junk}. One can 
readily check that $\frac{Q \bar Q}{H} \frac{Q \bar Q}{H} = \frac{Q \bar Q}{H}$. Thus, the fermion number operator 
$\frac{Q \bar Q}{H}$ has the eigevalues $p = 0, 1$ which implies the well-known Pauli's exclusion principle. The eigenstate 
$|\alpha \rangle_p$ of fermion number operator with eigenvalue $p =1$ is  a fermionic state whereas the eigenvalue $p= 0$ 
corresponds to a bosonic state.  The fermion number operator is a good quantum number operator because it is  Hermitian 
as well as commutes with the Hamiltonian.

Using the above algebraic relationships, we obtain
\begin{eqnarray}
\left(\frac{Q \bar Q}{H} \right) Q |\alpha\rangle_p &=& (p + 1) Q |\alpha \rangle_p, \nonumber\\ 
\left(\frac{Q \bar Q}{H} \right) \bar Q |\alpha\rangle_p &=& (p - 1) \bar Q |\alpha \rangle_p, \nonumber\\
\left(\frac{Q \bar Q}{H} \right) H |\alpha\rangle_p &=& p H |\alpha \rangle_p. 
\end{eqnarray}
These equations reflect the fact that the eigenstates $Q|\alpha \rangle_p$, $\bar Q|\alpha \rangle_p$ and $H|\alpha \rangle_p$
have the eigenvalues $(p+1)$, $(p-1)$ and $p$, respectively, with respect to the operator $(\frac{Q \bar Q}{H})$. If we identify 
the degree of a given form by the eigenvalue $p$, then we have 
{\it one-to-one} mapping between the conserved charges and de Rham cohomological operators:
\begin{eqnarray}
d \longleftrightarrow  Q, \qquad \delta \longleftrightarrow  \bar Q, \qquad \Delta \longleftrightarrow  H \equiv Q_w.
\end{eqnarray} 
It is now clear that the (lowering)raising  of the degree of a given form, by the operation of (co-)exterior derivative, is 
identified with the eigenvalue $p$ of the operator $(\frac{Q \bar Q}{H})$. As a consequence, the de Rham cohomological operators 
$(d, \delta, \Delta)$ and Hodge duality $(\star)$ operation find their physical realization in terms of conserved charges 
$(Q, \bar Q, Q_w \equiv H)$ and discrete symmetry transformations $(*)$, respectively and the present model provides a model 
for Hodge theory.

It is interesting to point out that there is yet another realization of the above differential operators in terms of the 
conserved charges if we define an eigenstate $|\beta \rangle_q$ with respect to an operator $\frac{\bar Q  Q}{H}$ 
such that $\frac{Q \bar Q}{H} |\beta \rangle_q = q |\beta \rangle_p$ where $q$ is an eigenvalue. This operator is also a 
fermion number operator and having eigenvalues $q = 0, 1$.  By exploiting the above algebra (\ref{34}), we obtain the followings
\begin{eqnarray}
\left(\frac{\bar Q Q}{H} \right) \bar Q |\beta \rangle_q &=& (q + 1) \bar Q|\beta \rangle_q, \nonumber\\ 
\left(\frac{\bar Q Q}{H} \right)  Q |\beta \rangle_q &=& (q - 1) Q |\beta \rangle_q, \nonumber\\
\left(\frac{\bar Q Q}{H} \right) H |\beta \rangle_q &=& q H|\beta \rangle_q, 
\end{eqnarray}
which show that the eigenstates $\bar Q|\beta \rangle_q$, $Q|\beta \rangle_q$ and $H |\beta \rangle_q$
have the eigenvalues $(q+1)$, $(q-1)$ and $q$, respectively, with respect to the operator 
$(\frac{\bar Q Q}{H})$. Keeping in mind the properties of  differential operators ($d, \delta, \Delta$), 
we have the following mapping:
\begin{eqnarray}
d \longleftrightarrow  \bar Q, \qquad \delta \longleftrightarrow Q, \qquad \Delta \longleftrightarrow  H \equiv Q_w.
\end{eqnarray}
It is clear that all the de Rham cohomological operators $(d, \delta, \Delta)$ find their physical analogue 
in terms of the conserved charges $(\bar Q, Q, Q_w)$. Thus, the ${\cal N} = 2$ supersymmetric quantum spin model  
turns out to be a model for Hodge theory.

\section{Conclusions}
\label{Sec. 7}
In our present investigation, we have discussed the various (fermionic and bosonic) continuous  as well as discrete symmetries of 
the ${\cal N} = 2$ supersymmetric quantum spin model. The continuous symmetries (and corresponding conserved charges) obey the 
standard algebra which is similar to the algebra followed by the de Rham cohomological operators of differential geometry.
Moreover, the interplay between fermionic and discrete symmetry transformations (in their operator form) provides the relations 
$s_2 = \pm * s_1*$ and $s_1 = \pm * s_2*$ which are similar to the relation $\delta = \pm \star d \star$. We have shown that there 
is a {\it one-to-one} mapping between the continuous symmetry transformations (and corresponding generators).  In other words, all 
the de Rham cohomological operators find their physical realization in terms of the symmetry properties of the  ${\cal N} = 2$ 
supersymmetric quantum spin model and the Hodge duality operation finds its physical meaning in terms of the the discrete symmetries. 
As a consequence,  the ${\cal N} = 2$ supersymmetric quantum spin model turns out to be a model for Hodge theory.

In the known literature, within the framework of BRST formalism, we have field-theoretic models which happen to be the models for 
Hodge theory in specific $D = 2p$-dimensions of spacetime where $p$ is the degree of a given differential form~\cite{rpm1,rpm2,rpm3,rpm4}. 
In these models, we have {\it two-to-one} mapping between the continuous symmetry transformations and de Rham cohomological operators 
where the continuous and nilpotent (anti-)BRST $(s_{(a)b})$, (anti-)co-BRST $(s_{(a)b})$ transformations and a unique bosonic transformation 
(and their corresponding generators) provide the physical realizations of the de Rham cohomological operators.

We note that, even though, the (anti-)BRST, (anti)-co-BRST and ${\cal N} = 2$ supersymmetric transformations are global and nilpotent 
of order two (i.e. $s^2_{(a)b} = s^2_{(a)d} = 0$ and  $s^2_1 = s^2_2 = 0$),  there are some glaring differences between them. The 
(anti-)BRST and  (anti)-co-BRST are anticommuting (i.e. $\{s_b, s_{ab}\} = \{s_d, s_{ad}\} = 0$) whereas ${\cal N} = 2$ supersymmetric 
transformations do not anticommute (i.e. $\{s_1, s_2 \} \ne 0$). On the other hand, the anticommutators 
$\{s_b, s_d\} = - \{s_{ab}, s_{ad}\} = s_\omega$   lead to a unique bosonic symmetry $s_\omega$. Similarly, the anticommutator 
$\{s_1, s_2\} = s_w$ of supersymmetric transformations $s_1$ and $s_2$ also defines a bosonic symmetry transformation. But, $s_w$ 
is different from $s_\omega$ in the sense that the former transformation produces the time translation (i.e. $s_w \chi \approx \dot \chi$) 
and, thus, Hamiltonian is the  generator of $s_w$ whereas the latter symmetry $s_\omega$ does not emerge form Hamiltonian.

In view of our present work, it would be an interesting piece of work to study the quantum spherical model where the  spin
variables $S_{\bf r}$ are subjective to a spherical constraint $\displaystyle\sum_{\bf r} S^2_{\bf r} - N =0.$ Here $N$ is 
the total number of lattice sites. Furthermore, the supersymmetrization of quantum spherical model with local 
symmetry~\cite{gomes} is also an interesting work. We point out that our idea can also be applied to the ${\cal N } = 2,4$ 
supersymmetric field theories having local gauge symmetries. These works are under investigation and we shall report elsewhere.

\acknowledgments{RK would like to thank UGC, Government of India, New Delhi for financial support under the PDFSS scheme.
RK and AS are very much thankful to the Head, Department of Physics, Banaras Hindu University, Varanasi for inviting us
and taking care of hospitality during the visit where this work has been started.}

\renewcommand{\theequation}{A.\arabic{equation}}    
  \setcounter{equation}{0}  

\appendix

\section{Appendix A: On-shell supersymmetric transformations}
\label{Appendix A}
The Lagrangian (\ref{lag2}) can be simplified by eliminating an auxiliary variable $F_{\bf r}$. Using the equation of motion for 
$F_{\bf r}$ (cf. (\ref{25})) in (\ref{lag2}), we obtain 
\begin{eqnarray}
L_0 &=& \frac{1}{2g}\,\sum_{\bf r} {\dot S}_{\bf r} ^2 + \frac{i}{\sqrt{g}} \sum_{\bf r} \bar \Psi_{\bf r}  \, \dot \Psi_{\bf r}  
- \sum_{{\bf r},{\bf r}'} U_{{\bf r}, {\bf r}'}\, \bar \Psi_{\bf r} \,\Psi_{{\bf r}'} \nonumber\\
&& -\, \frac{1}{2} \sum_{{\bf r},{\bf r}'} J_{{\bf r}, {\bf r}'} \,S_{\bf r} \,S_{{\bf r}'},
 \label{39}
\end{eqnarray}
where we have defined 
\begin{eqnarray}
\sum_{\bf s} U_{{\bf r}, {\bf s}} U_{{\bf s}, {\bf r}'} \equiv J_{{\bf r},{\bf r}'}.
\end{eqnarray}
It is to be noted that the above Lagrangian is coincide with the Lagrangian given in~\cite{gomes1}. Similarly, one can eliminate 
$F_{\bf r}$ from the  supersymmetric transformations (\ref{20}) and obtain the following on-shell nilpotent 
transformations, namely; 
\begin{eqnarray}
&& s_1 S_{\bf r} = i\, \sqrt{g} \,\Psi_{\bf r},   \qquad s_1 \Psi_{\bf r} = 0, \nonumber\\
&& s_1 \bar \Psi_{\bf r} = - \big(\dot S_{\bf r} + i \sqrt{g}\,\sum_{{\bf r}'} U_{{\bf r}, {\bf r}'} S_{{\bf r}'} \big), \nonumber\\
&& \nonumber\\
&& s_2 S_{\bf r} = i\, \sqrt{g} \bar \Psi_{\bf r}, \qquad s_2 \bar  \Psi_r = 0, \nonumber\\
 && s_2  \Psi_{\bf r} = -  \big(\dot S_{\bf r} -  i \sqrt{g}\,\sum_{{\bf r}'} U_{{\bf r}, {\bf r}'} S_{{\bf r}'}\big), 
\end{eqnarray}
where the nilpotency of $s_1$ and $s_2$ can be shown in a straightforward by using the Euler-Lagrange equations of motion 
derived from (\ref {39}).


\begin{thebibliography}{99}
\bibitem{huang}  K. Huang, {\it Statistical Mechanics} (John Wiley and Sons, New York, 1963).
\bibitem{gr}     W. Greiner and L. Neise, H. Stocker, {\it The Models of Ising and Heisenberg} 
                 (Springer, New York, 1995).
\bibitem{jo}     G. S. Joyce, {\it Phys. Rev.} {\bf 155},  478 (1967).
\bibitem{aa}     A. A. Zvyagin and G. A. Skorobagat'ko, {\it Phys. Rev. B} {\bf 73}, 024427 (2006).

\bibitem{dm}     D. V. Dmitriev, V. Ya. Krivnov, A. A. Ovchinnikov, 
                 {\it J. of Exp. and Theor. Phys.} {\bf 88}, 138 (1999).

\bibitem{chakr}  S. Chakravarty, B. I. Halperin  and D. R. Nelson, {\it Phys. Rev. B}  {\bf 39}, 2344 (1989). 
\bibitem{sachdev} S. Sachdev and J. Ye, {\it Phys. Rev. Lett.}  {\bf 69}, 2411 (1992).
\bibitem{sachdev1} J. Ye, S. Sachdev and N. Read, {\it Phys. Rev. Lett.} {\bf 70}, 4011 (1993).
\bibitem{Sousa}  J. R. De Sousa,  {\it Physica A: Staistical Mechanics and Its Applications} {\bf 259}, 138 (1998)
\bibitem{ig}     I. Glasser, J. I. Cirac, G. Sierra, A. E. B. Nielson, {\it Nucl. Phys. B} {\bf 886}, 63 (2014)
\bibitem{joy}    G. S. Joyce, {\it Phys. Rev.}, {\bf 146} 349 (1966).
\bibitem{kac}    T. H. Berlin and M. Kac, {\it Phys. Rev.} {\bf 86},  821 (1952).
\bibitem{joy1}   G. S. Joyce, {\it Phase Transitions and Critical Phenomena}, \\edited
                 by C. Domb and M. S. Green, Vol. {\bf 2} (Academic Press, New York, 1972).
\bibitem{joy3}   G. S. Joyce {\it Phys. Rev. Lett.} {\bf 19}, 581 (1967).
\bibitem{sch}      T. Schneider,  E. Stoll and H. Beck, {\it Physica A}  {\bf 79}, 201 (1975).
\bibitem{pt}       N. M. Plakida and N. S. Tonchey, {\it Physica A} {\bf 136}, 176 (1986).
\bibitem{vojta}    T. Vojta, {\it Phys. Rev.  B}  {\bf 53}, 710 (1996).
\bibitem{singh}    P. Shukla and S. Singh, {\it Phys. Lett. A} {\bf 81}, 477 (1981).



\bibitem{tn}     T. M. Nieuwenhuizen, {\it Phys. Rev. Lett.} {\bf 74}, 4293 (1995).
\bibitem{gomes}  P. R. S. Gomes and  P. F. Bienzobaz, {\it Phys, Rev. E}  {\bf 91}, 022122 (2015).
\bibitem{gomes1} P. R. S. Gomes,  P. F. Bienzobaz and M. Gomes, {\it Phys. Rev. E} {\bf 85}, 061109 (2012).
\bibitem{das}    A. Das and M. Kaku, {\it Phys. Rev. D} {\bf 18}, 4540 (1978).


\bibitem{umezawa} H. Matsumoto, M. Nakahara, Y. Nakano  and H. Umezawa, {\it Phys. Rev.} D, {\bf 29}, 2838 (1984).
\bibitem{das1}  A. Das,  A. Khare and V. S. Mathur, {\it Phys. Lett. B} {\bf 181}, 299 (1986).
\bibitem{vafa}  S. Cecotti and C. Vafa, {\it Commun. Math. Phys.} {\bf 157}, 139 (1993).
\bibitem{rpm1}  R. P. Malik, {\it Int. J. Mod. Phys. A} {\bf 22}, 3521 (2007).
\bibitem{rpm2}  S. Gupta and R. P. Malik, {\it Eur. Phys. J. C} {\bf 58}, 517 (2008). 
\bibitem{rpm3}  R. Kumar, S. Krishna, A. Shukla and R. P. Malik, {\it Eur. Phys. J. C} {\bf 72}, 1980 (2012).
\bibitem{rpm4}  R. Kumar, S. Krishna, A. Shukla and R. P. Malik, 
                {\it Int. J. Mod. Phys. A} {\bf 29},  1450135 (2014). 
\bibitem{rpm5}  S. Gupta and R. P. Malik, {\it Eur. Phys. J. C}  {\bf 68}, 325 (2010).
\bibitem{ra}    R. Kumar and A. Shukla, {\it Europhys. Lett.} {\bf 115}, 21003 (2016). 
\bibitem{rm}    R. Kumar and R. P. Malik, {\it Eur. Phys. J.  C} {\bf 73}, 2514 (2014).
\bibitem{rm1}   R. Kumar and R. P. Malik, {\it Europhys. Lett.} {\bf 98}, 11002 (2012).
\bibitem{kam}   S. Krishna, A. Shukla and R. P. Malik, {\it Ann. of Phys.} {\bf 351}, 558 (2014).
\bibitem{as}    A. Shukla, {\it Adv. in High Energy Phys.} {\bf 2017}, 1403937  (2017).
\bibitem{egh}   T. Eguchi, P. B. Gilkey and A. Hanson, {\it Phys. Rep.}, {\bf 66} (1980) 213.
\bibitem{vh}    J. W. van Holten, {\it Phys. Rev. Lett.} {\bf 64}, 2863 (1990).
\bibitem{dght}  S. Deser, A. Gomberoff, M. Henneaux and C. Teitelboim,  {\it Phys. Lett. B}  {\bf 400}, 80 (1997).
\bibitem{smm}   S. Mukhi and N. Mukunda, {\it Introduction to Topology, Differential 
                Geometry and Group Theory for Physicists} (Wiley Eastern Private Limited, New Delhi, 1990). 
                
\bibitem{salo} P. Salomonson, J. W. and van Holten, {\it Nucl. Phys. B} {\bf 196}, 509
                (1982). 
\bibitem{junk} G. Junker, {\it Supersymmetric Methods in Quantum and Statistical Physics},  
               (Springer-Verlag Berlin Heidelberg, 1996).                 
\end{thebibliography}
\end{document}